\documentclass[aip,jmp,twocolumn,amsmath,amssymb,showpacs]{revtex4-1}
\usepackage{amssymb}
\usepackage{epsfig}
\usepackage{amsmath}
\usepackage{times}

\begin{document}

\title{Fluctuation of Dynamical Robustness in a Networked Oscillators System}

\author{Wenwen Huang}
\affiliation{Department of Physics, East China Normal University,
Shanghai, 200062, P. R. China}

\author{Xiyun Zhang}
\affiliation{Department of Physics, East China Normal University,
Shanghai, 200062, P. R. China}

\author{Xin Hu}
\affiliation{Department of Physics, East China Normal University,
Shanghai, 200062, P. R. China}

\author{Zonghua Liu}
\affiliation{Department of Physics, East China Normal University,
Shanghai, 200062, P. R. China}

\author{Shuguang Guan}
\email{guanshuguang@hotmail.com} \affiliation{Department of Physics,
East China Normal University, Shanghai, 200062, P. R. China}

\date{\today}

\begin{abstract}
In this work, we study the dynamical robustness in a system consisting of both active and inactive oscillators.
We analytically show that the dynamical robustness of such system is determined by the cross link density between active and inactive subpopulations, which depends on the specific process of inactivation. It is the multi-valued dependence of the cross link density on the control parameter, i.e., the ratio of inactive oscillators in the system, that leads to the fluctuation of the critical points.   We further investigate how different network topologies and inactivation strategies affect the fluctuation. Our results explain why the fluctuation is more obvious in heterogeneous networks than in homogeneous ones, and why the low-degree nodes are crucial in terms of dynamical robustness. The analytical results are supported by numerical verifications.

\end{abstract}

\pacs{89.75.Hc,05.45.Xt,89.20.-a}

\keywords{complex network, synchronization, dynamical robustness}

\maketitle

\begin{quotation}
Robustness is an important aspect of the nature of a system. In the past years, topological robustness has been deep studied,
but the dynamical robustness did not catch much attention. In this paper, we study the dynamical robustness of a system comprised of Stuart-Landau oscillators. The oscillators are divided into two kinds: active and inactive. When the ratio of inactive oscillators increase, the status of the system will from oscillation change to amplitude death and the critical point can be used to measure the Robustness of the sysrem. Using numerical simulation and theoretical analysis, we find that the cross link density between active and inactive subpopulations determins the dynamical robustness of the system.

\end{quotation}

\section{Introduction}

Dynamical systems in nature and human labs usually comprise a large number of  interacting individual elements, such as synchronizing fireflies \cite{buck1988synchronous},  neurons in human brain \cite{izhikevich2007dynamical} , cardiac pacemaker cells \cite{glass2001synchronization}, power grids \cite{filatrella2008analysis,Rohden2012}, and Josephson junction arrays \cite{wiesenfeld1996synchronization},  just to name a few, which can be modeled by networked oscillators \cite{pikovsky2003synchronization}.
One important issue in the study of such dynamical systems is the robustness, i.e., the ability  to maintain basic structure and function under attacks or disfunctions \cite{Sergey2010catastrophic,Duncan2000percolation,Vespignani2010,morino2011robustness,Tanaka2012}. It has been shown that the huge blackouts, which inevitably cause tremendous economic loss, are related to the cascading failure of power-grids \cite{Sergey2010catastrophic}.
Such structure robustness involving network connectedness
has been intensively investigated in previous works  \cite{Sergey2010catastrophic,Duncan2000percolation,Vespignani2010}. On the other hand,
networked systems typically carry dynamics, for example,
the circadian rhythms of mammals, the synchronization of
cardiac cells, and spatio-temporal patterns in brain \cite{izhikevich2007dynamical,glass2001synchronization,izhikevich2007dynamical}.
Of equal importance is the dynamical robustness, i.e., the ability  of a networked system to maintain its normal dynamical activity when the topology or the local dynamics of network components are subjected to changes.

In Ref. \cite{Daido2004},
Daido et al. investigated the dynamical robustness of a globally coupled system which simultaneously consists of active and inactive oscillators. With the increase of the ratio of inactive oscillators,  the dynamical activity in the system decreases until finally it totally vanishes at certain critical point. Such phenomenon is termed as aging transition which characterizes the dynamical robustness of such networked systems \cite{Daido2004,morino2011robustness}.
Recently, Tanaka et al. further studied this phenomenon in various complex network topologies \cite{Tanaka2012}. Surprisingly, it is found that the low-degree nodes, rather than the hubs, play a dominant role in terms of dynamical robustness. This finding uncovered the great differences between dynamical robustness and structure robustness.

 For such dynamical systems, we always expect the dynamical robustness of them could be controllable. While sometimes we wish a strong robust system to maintain its functionality when suffering attacks, sometimes we prefer a fragile system because it is harmful and we need to destroy it. Tanaka's work reported that coupled oscillators network with smaller mean degree appears more robust, which give us a hint that we could increase the dynamical robustness of the system by reducing connections between elements. However, in this paper, we report the fluctuation of dynamical robustness in the above mentioned networked system. We find that even the network topology and the coupling strength in the dynamical system are fixed, there is still a fluctuation for the critical point of aging transition measured by the ratio of inactive oscillators in the system, which characterizes the dynamical robustness of the system. Moreover, the fluctuation could be large enough that totally changes the robustness property of the system. So it is important to understand the fluctuation for the potential applications of dynamical robustness. In addition, this fluctuation occurs more obviously in heterogeneous networks than in homogeneous ones. By applying mean-field approximation and linear stability analysis, we show that
the cross link density between subpopulations of active and inactive oscillators plays a dominant role. In principle, different inactivation realizations lead to different link densities, which finally result in fluctuations of the critical points. Based on this analysis,  we can explain  why the fluctuations in heterogeneous networks are severer than that in homogeneous networks. Our results are also helpful to understand the crucial role of the low-degree nodes in determining the dynamical robustness of networked oscillators reported in Ref. \cite{Tanaka2012}.
The rest of this paper is organized as follows: section II introduces the dynamical model; theoretical analyses and numerical verifications are presented in section III; finally, conclusions are presented in section IV.

\section{Model}

In this work, we investigate a dynamical model of networked oscillators. It  has following features: (1) the dynamical system is described by coupled oscillators; (2) there are two types of oscillators in the system, i.e., active and inactive, representing oscillating state and non-oscillating state, respectively; (3) the coupling among oscillators forms a complex network.  The general form of this model can be written as:
\begin{equation}\label{eq:fcp}
\dot{\mathbf{x}_{j}}=\mathbf{F}(\mathbf{x}_{j})+\sigma\sum_{k=1}^{N}c_{jk}(\mathbf{x}_{k}-\mathbf{x}_{j}).
\end{equation}
Here, $j=1, \cdots, N$ is the index of oscillator or node. $\mathbf{x}$ is the state vector describing the dynamics of oscillators. The first term at the RHS of Eq. (\ref{eq:fcp}) describes the local dynamics of an oscillator, and the second term is the diffusive coupling that denotes interactions among different oscillators. $\sigma$ is the coupling strength.
$c_{jk}$ is the element in the adjacent matrix of coupling network, which equals to 1 if nodes $j$ and $k$ are connected, and 0 otherwise. It should be pointed out that
similar models have been studied previously.  For example, aging transition in such models was investigated in fully coupled network in Ref. \cite{Daido2004}, and later in regular ring topology in Refs \cite{Daido2008,Daido2011b}. Recently, it has been extended to different complex topologies \cite{morino2011robustness,Tanaka2012}.

In the present work, we mainly choose the Stuart-Landau (SL)  oscillators as the local dynamics. Specifically, the networked SL oscillators system can be described by the following coupled ODEs:
\begin{equation}\label{eq:slcp}
\dot{z_{j}}=(\alpha_{j}+i\Omega_{j}-|z_{j}|^{2})z_{j}+\sigma\sum_{k=1}^{N}c_{jk}(z_{k}-z_{j}),
\end{equation}
where $z_{j}$ and $\Omega_{j}$ are the complex amplitude and the inherent frequency of the $j$th SL oscillator, respectively.
$\alpha_{j}$ is the control parameter denoting the distance from a Hopf bifurcation point.
When $\alpha_{j}>0$, the oscillator is a limit cycle with an amplitude $\sqrt{\alpha_{j}}$. However, it settles down to a fixed point when $\alpha_{j}<0$. Thus generally the oscillator is active when $\alpha_{j}>0$ and inactive when $\alpha_{j}<0$.
The oscillator will lose its activity as its $\alpha$ value converts from positive to negative. This can be used to model the two distinct dynamical states of oscillators. We define parameter $\rho$ as the ratio of inactive oscillators in the network. Reasonably, the global activity of the networked system can be characterized by the normalized order parameter $Q$, defined as $Q=|Z(\rho)|/|Z(0)|$ with $Z=N^{-1}\sum_{j=1}^{N}z_{j}$. As $\rho$ increases to a critical value $\rho_c$, the networked system will totally loses its global activity until $Q=0$, i.e., an aging transition occurs \cite{Daido2004}. Because the ratio $\rho_c$ is the largest
ratio with which the dynamical system can maintain certain activity, it measures the dynamical robustness of this networked system. The larger the $\rho_c$, the better the dynamical robustness.

\begin{figure}
  \epsfig{figure=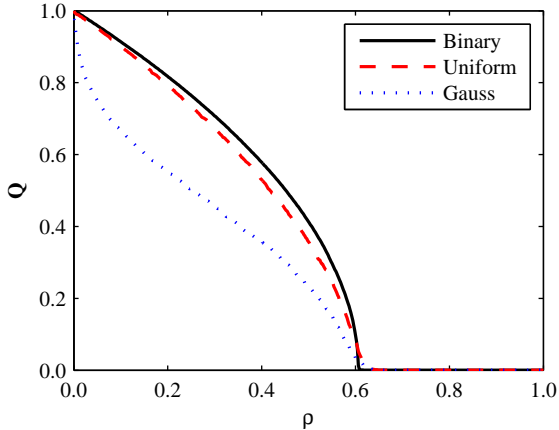,width=1.0\linewidth}
  \caption{(colour online) The aging transitions of three typical $\alpha$ distributions on a small-world network. For binary distribution, $a=1, b=1$; for uniform distribution and Gaussian distribution, the mean $\alpha$ varies from 3 to -3. The uniformly distributed $\alpha$ varies in a range of length 6, and $\alpha$ of Gaussian distribution has the standard variance 1. System size $N=500$, the mean degree $\langle K \rangle =50$, $\sigma=0.1$, $\Omega=3$. Average over $100$ times.} The same results have been obtained on scale-free networks and regular lattices.   \label{fig:AlphaD}
\end{figure}

For simplicity, we will not consider the case in which the frequencies of oscillators have a distribution. Throughout this paper, we set $\Omega_{1}=\Omega_{2}=\cdots=\Omega_{N}=\Omega$.
Since the model simultaneously contains both active and inactive local dynamics characterized by  positive and negative $\alpha$, respectively,  we  first examine how the configuration of $\alpha$ values affect the global dynamics. To this end, we consider three typical distributions of $\alpha$. The first and the simplest case is $\alpha_{j}=a>0$ for all active oscillators, while $\alpha_{j}=-b<0$ for all inactive ones, i.e., $\alpha_{j}$ is binary.  In the second and the third cases, $\alpha_{j}$ satisfies uniform distribution and Gaussian distribution, respectively, i.e., $\alpha_{j}$ varies in a range. The ratio of the inactive oscillators can be tuned by gradually changing the mean of the distribution from positive to negative.
Fig. \ref{fig:AlphaD} illustrates the aging transition for the three types of  $\alpha$ settings. Interestingly,
it is found that although the values of $\alpha$ satisfy different distributions, what really determines the transition point is the the ratio of the inactive oscillators in the system. Therefore,
we can always  equivalently simplify the distribution of $\alpha$ into binary case by coarse-graining process.  Without loss of generality, in the following we only consider the situation where $\alpha$ takes binary values.

Besides SL oscillators, we also consider the following  networked R\"{o}ssler oscillators in this paper:
\begin{equation}
  \label{eq:rosslercp}
  \begin{array}{c}
  \dot{x_j}=-y_j-z_j+\sigma\sum\limits^{N}_{j=1}{c_{jk}(x_k-x_j)},         \\
  \dot{y_j}=x_j+c_jy_j+\sigma\sum\limits^{N}_{j=1}{c_{jk}(y_k-y_j)},       \\
  \dot{z_j}=d_j+z_j(x_j-e_j)+\sigma\sum\limits^{N}_{j=1}{c_{jk}(z_k-z_j)}, \\
  \end{array}
\end{equation}
where $x,y,z$ are the state variables of R\"{o}ssler oscillator, and $c,d,e$ are parameters. In this system, there are also two types of oscillators by choosing different parameters, i.e,
$c=d=0.2, e=1$ for active oscillators  and $c=d=-0.2, e=2.5$ for inactive oscillators. To quantify the global activity, the order parameter can be defined as $Q=\sqrt{\langle (\mathbf{x}_{c}-\langle\mathbf{x}_{c}\rangle)^2 \rangle}$ , where $\mathbf{x}_{c}= \sum_{j=1}^{N}(x_{j},y_{j},z_{j})/N$ is the centroid and the bracket means the long time average.
Numerical results thoughout this work were obtained for random initial conditions by means of the fourth order Runge-Kutta method with time step 0.01.

\begin{figure}
\epsfig{figure=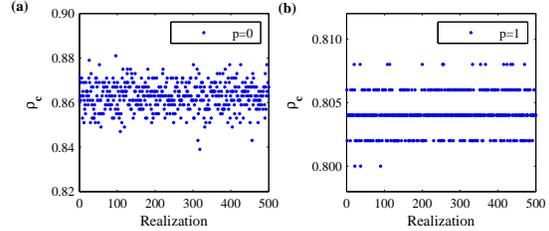,width=1.0\linewidth}
\caption{(colour online) Fluctuations of dynamical robustness characterized by the transition points $\rho_c$ in networked system of SL oscillators.  $p$ is the parameter characterizing the homogeneity/heterogeneity of the network. As $p$ goes from 1 to 0, the network continuously changes from  heterogeneous to homogeneous. See Appendix for detail. $N=500$, $\langle K \rangle=50$, $\sigma=0.1$, $a=2$,  $b=1$, $\Omega=3$. }\label{fig:fluctuation}
\end{figure}

\begin{figure}
  \epsfig{figure=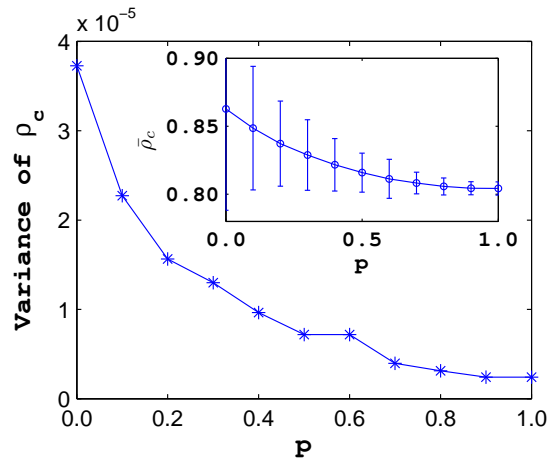,width=1.0\linewidth}
  \caption{(colour online) Variance of $\rho_c$ versus parameter $p$ in networked system of SL oscillators. Inset is the mean value of $\rho_c$ for 500 realizations. $N=500$, $\sigma=0.1$, $a=2,b=1,\Omega=3$. }  \label{fig:Deltarhoc}
\end{figure}

\section{Analysis and results}

Recently, in our study we find an interesting phenomenon, i.e., even for fixed network topology and coupling strength, the critical point $\rho_c$ has fluctuations depending on specific network realizations. Typical examples are shown in Fig. \ref{fig:fluctuation}.
In our study, in order to conveniently compare different network topologies, we propose a simple but effective method to continuously change a heterogeneous network into a homogeneous one. The detail of the algorithm is explained in Appendix. As shown in Fig. \ref{fig:fluctuation}, fluctuations of critical point $\rho_c$ are observed in both heterogeneous and homogeneous networks. To quantify this feature, in Fig. \ref{fig:Deltarhoc}, we plot the variance of $\rho_c$ when the network topology continuously changes from heterogeneous to homogeneous with the increase of rewinding probability $p$. It is found that this phenomenon is more obvious in heterogeneous networks.
In the following, we present both theoretical analyses and numerical verifications to understand this phenomenon.
We notice that some non-mean-field methods, e.g., in Ref. \cite{Restrepo2005},  have been proposed to deal with synchronization of Kuramoto-type phase oscillators on complex networks. However,  a general theoretical framework to treat networked periodical and chaotic oscillators involving amplitudes is still not available. Basically, the analyses in the present study are based on mean-field approximation.

\begin{figure}
\epsfig{figure=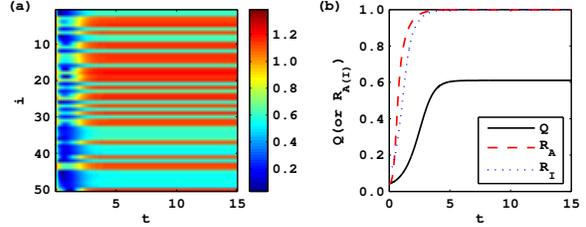,width=1.0\linewidth}
\caption{(colour online) (a) The evolution of amplitude $|z_{i}(t)|$ in a homogeneous network ($p=1$) with $N=50$.  $\rho=0.5 < \rho_c$, $\langle K \rangle =10, \sigma=0.1, a=2, b=1, \Omega=3$. The colour bar denotes the magnitude of amplitude.  (b) The global order parameter ($Q$), and the order parameters of active ($R_A$) and inactive ($R_I$) subpopulations in a heterogeneous network ($p=0$). It is shown that the oscillators inside either subpopulation achieve synchronization, but the whole system do not achieve global synchronization. Here, $R_{A(I)}=|\sum_{j\in S_A(S_I)}z_j|/\sum_{j\in S_A(S_I)}|z_j|$, $N=500$, $\langle K \rangle =30, \sigma=0.1,  a=2, b=1, \Omega=3$.}
\label{fig:Sync}
\end{figure}

\begin{figure*}
\epsfig{figure=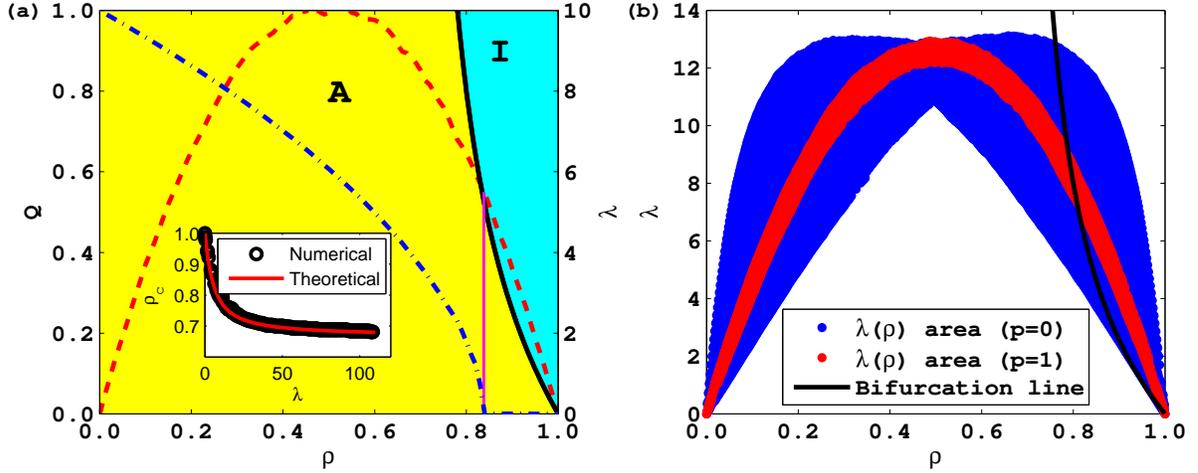,width=1.0\linewidth}
\caption{(colour online) The $\rho$-$\lambda$ parameter panel. (a) The solid black line is the bifurcation curve defined by Eq. (\ref{rhoclambda}); the red dash line is the curve $\lambda(\rho)$ for a specific realization of inactivation ($p=1$); and the blue dot dash line corresponds to the order parameter. Inset compares the theoretical result of $\rho_c$ with that of numerical experiments. (b) The solid black line is the same as in (a); the blue area shows the variation area of $\lambda(\rho)$ curve in heterogeneous network ($p=0$) while the red area corresponds to that in homogeneous network ($p=1$). Other parameters are the same in both (a) and (b): $N=500, \langle K \rangle =50, \sigma=0.1,  a=2, b=1, \Omega=3$.}
\label{fig:panel}
\end{figure*}

\subsection{Linear stability analysis}

We start from Eq. (\ref{eq:slcp}).
When $\rho=0$, i.e., the system contains only identical active oscillators. In this case, it will easily achieve global synchronization as the coupling strength increases.
When we randomly choose some nodes and change the oscillators into inactive states, i.e., the proportion of inactive nodes $\rho>0$, the system simultaneously contains both active and inactive oscillators. Such a process is called random inactivation. Throughout this paper, it is adopted by default unless otherwise stated.
Usually, the non-trivial aging transition would only occur when the coupling strength is large enough \cite{Daido2004}, so before the transition all the active oscillators already well synchronize into a cluster, and so does the inactive oscillators. This has been numerically verified and one such example is illustrated in Fig. \ref{fig:Sync}. Although the synchronization is not rigorous, it is good enough for us to
approximately reduce the high-dimensional system into two identical clusters of oscillators, denoted as $S_A$ and $S_I$ for active and inactive oscillators, respectively.
We use $A$ and $I$ to represent the dynamics of active and inactive subpopulations, respectively, as treated in Ref. \cite{Daido2004}. Then the dynamics of the networked system, i.e., Eq. (\ref{eq:slcp}), can be
written as:
\begin{eqnarray}\label{eq:slaicp}
\dot{A}=(a+i\Omega-|A|^{2})A+\sigma\mu_{j}K_{j}(I-A),  \\
\dot{I}=(-b+i\Omega-|I|^{2})I+\sigma(1-\mu_{j})K_{j}(A-I).
\end{eqnarray}
Here, $K_{j}$ means the degree of oscillator $j$, and $\mu_{j}$ is the proportion of inactive oscillator among the neighbors of oscillator $j$.
Summing the equations of all active and inactive oscillators, we obtain
\begin{eqnarray}
  \label{eq:sumslaicp1}
  \dot{A}=(a+i\Omega-|A|^{2})A+\frac{\sigma(I-A)}{(1-\rho)N}\sum_{j\in S_A}\mu_{j}K_{j}, \\
  \label{eq:sumslaicp2}
  \dot{I}=(-b+i\Omega-|I|^{2})I+\frac{\sigma(A-I)}{\rho N}\sum_{j\in S_I}(1-\mu_{j})K_{j}.
\end{eqnarray}
In the above equations,
the summing terms at the RHS, i.e., $\sum_{j\in S_A}\mu_{j}K_{j}$ and $\sum_{j\in S_I}(1-\mu_{j})K_{j}$, both represent the total number of cross links, denoted by $L$, between the subpopulations of active oscillators and inactive ones. Actually they are the same thing in different ways. To avoid the influence of network size, we define the density of cross links as $\lambda=L/N=\sum_{j\in S_A}\mu_{j}K_{j}/N=\sum_{j\in S_I}(1-\mu_{j})K_{j}/N$.
Then Eqs. (\ref{eq:sumslaicp1}) and (\ref{eq:sumslaicp2}) become
\begin{eqnarray}
  \label{eq:lambdaslaicp1}
  \dot{A}=(a+i\Omega-|A|^{2})A+\frac{\sigma\lambda(I-A)}{(1-\rho)}, \\
  \label{eq:lambdaslaicp2}
  \dot{I}=(-b+i\Omega-|I|^{2})I+\frac{\sigma\lambda(A-I)}{\rho}.
\end{eqnarray}
From these equations, we immediately find that the network topology actually affects the global dynamics through parameter $\lambda$. It is the cross link density between active and inactive subpopulations that plays a dominant role in this networked system.
Now we analytically study how the dynamical robustness of such system can be affected by parameter $\lambda$.
With the increase of control parameter $\rho$, the dynamics of the networked system will gradually lose global activity. This process can be characterized by $Q \rightarrow 0$ when $\rho\rightarrow \rho_c$. At the transition point, the networked system loses its stability and in the mean time the trivial fixed point $z_{0}=(A,I)=(0,0)$ becomes stable. By a linear stability analysis, we obtain the critical point $\rho_c$ as:
\begin{equation}\label{rhoclambda}
\rho_{c}=\frac{ab-\sigma (a+b)\lambda+[ab+\sigma (a+b)\lambda]\sqrt{1-\beta}}{2ab},
\end{equation}
with $\beta=4\sigma ab^2\lambda/[ab+\sigma(a+b)\lambda]^2$. For typical dense complex network, $\lambda \gg 1$. Therefore, $\beta$ is a small quantity.
We apply  Taylor expansion to $\sqrt{1-\beta}$ in Eq. (\ref{rhoclambda}) and keep the linear term. It finally becomes
\begin{equation}\label{rhoclambdataylor}
\rho_{c}=1-\frac{\sigma b}{ab/\lambda+\sigma(a+b)}.
\end{equation}

\subsection{Analysis to fluctuation of critical point }

Eq. (\ref{rhoclambdataylor}) shows that $\rho_c$ only has functional relation with $\lambda$ given that the parameters of the local dynamics and the coupling strength are fixed.  Apparently, there is a maximal value $\rho_c^{max}=1$ when $\lambda \rightarrow 0$. With the increase of $\lambda$, $\rho_c$ will monotonically decrease. When $\lambda \rightarrow \infty $, $\rho_c$ approaches the minima $\rho_c^{min}=\frac{a}{a+b}$, as shown in Fig. \ref{fig:panel}.
This result provides us insight that the dynamical robustness of such system is determined by the density of cross links $\lambda$. Physically, this can be reasonably understood. Since there are both active and inactive local states in the network, the interaction or influence between the two subpopulations are crucial for the global activity. The density of cross links $\lambda$ just characterizes this interaction. For example, when it is large, there exists strong interaction between the two subpopulations of active and inactive oscillators. As a result, a small critical value $\rho_c$ can be expected.

Now we explain why there exists fluctuation of critical point $\rho_c$ even when the network topology is fixed.
Let us analyze the phase diagram  on the parameter panel of $\rho$-$\lambda$. As shown in Fig. \ref{fig:panel} (a),
Eq. (\ref{rhoclambda}) actually defines the curve of bifurcation, i.e., the boundary of two areas corresponding to distinct dynamical phases of the system. In the upper right area, the networked system is in the quenching state losing global activity; while in the bottom left area, it is in active state oscillating to some extent.
The active state loses its stability when the system passes through the curve Eq. (\ref{rhoclambda}).
For an inactivation process, i.e,. in a specific realization, $\lambda$ is a function of $\rho$.
Actually, $\lambda$ is a unimodal function satisfying $\lambda(0)=\lambda(1)=0$, as shown in
Fig. \ref{fig:panel} (a). The curve $\lambda(\rho)$ intersects
the bifurcation curve Eq. (\ref{rhoclambda}), and the crosspoint just determines the critical point $\rho_c$.
The key point here is that  $\lambda$ is a multi-valued function of $\rho$ even when the network topology is fixed.
For different numerical realizations, i.e., in different specific inactivation processes, we will get different curves
$\lambda(\rho)$, which lead to different crosspoints in the bifurcation curve Eq. (\ref{rhoclambda}). This is the origin of fluctuation of critical point $\rho_c$ observed in our study.

Naturally, one may want to know which inactivation process
will give the maximal $\rho_c$ or the minimal $\rho_c$, which directly limits the fluctuation scope of $\rho_c$ given the network topology.
Interestingly, this problem can be mapped into the ground state problem of anti-ferromagnetic Ising model or the MAX-CUT problem in combinatorial optimization \cite{Garey1979,Laurent1979,Zhou2005}. It has been proved that the solution of these problems in general networks is a NP-Complete problem, i.e., we cannot find an effective algorithm to determinate it in polynomial time \cite{Garey1979}.
Since analytical method is not available to get the fluctuation scope of $\rho_c$, we turn to numerical study.
To this end, we numerically plot the possible curves $\lambda(\rho)$ on the parameter panel of $\rho$-$\lambda$.
The two crosspoints that define the largest range on the axis of $\rho$ roughly gives the fluctuation range of $\rho_c$.
As shown in Fig. \ref{fig:panel}(b), it is found that  $\lambda(\rho)$ for a heterogeneous network has a much larger fluctuation area compared to a homogeneous network. Therefore,  more obvious fluctuation of $\rho_c$ is observed in the former case. Physically, this result can be heuristically understood.
We know that the fluctuation of $\rho_c$ is caused by the multi-valued $\lambda(\rho)$, which  depends on specific inactivation processes.
Obviously,  an inactivation process is  more sensitive to the topology in heterogeneous networks
than in homogeneous ones because there are huge differences among the node degrees in the former case. This leads to the fluctuation of $\lambda(\rho)$
to a greater degree, which finally contributes to the fluctuation of the critical point $\rho_c$.

\begin{figure}
  \epsfig{figure=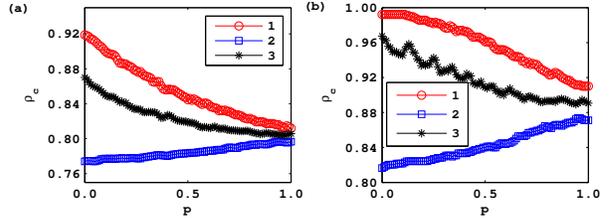,width=1.0\linewidth}
  \caption{(colour online) Dynamical robustness characterized by $\rho_c$ under three typical strategies of inactivation. (a)  Networked system of SL oscillators. $\sigma=0.1$, $a=2$, $b=1$, $\Omega=3$. (b) Networked system of R\"ossler oscillators. $c=d=0.2, e=1$ for active oscillators and $c=d=-0.2, e=2.5$ for inactive ones. $\sigma=0.002$. Other parameters are the same for (a) and (b):  $N=500$, $\langle K \rangle=50$. }
  \label{fig:Strategy}
\end{figure}

\begin{figure}
  \epsfig{figure=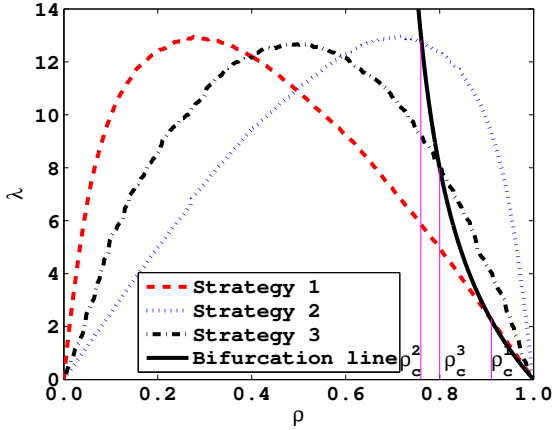,width=1.0\linewidth}
  \caption{(colour online) Identifying the critical points
  under  three typical strategies of inactivation in networked
   system of SL oscillators. The meanings of curves are the same as in Fig. \ref{fig:panel}. $p=1$, and other parameters are the same as in Fig. \ref{fig:Strategy}. }  \label{fig:Strapanel}
\end{figure}

As one application of the above theory, we now consider the case of targeted inactivation rather than random activation in network.
In Ref. \cite{Tanaka2012},
It has been reported that under targeted inactivation the dynamical robustness of the system depends more on the low-degree nodes rather than  the hubs. This important finding reveals the crucial role of the low-degree nodes in the context of dynamical robustness. Based on our analytical treatment, we can provide an explaination here.  In our study,
we apply three typical strategies of inactivation: (1) Inactivation goes from the node with the maximal degree to the one with the minimal degree; (2) Inactivation takes the inverse order of (1); (3) Random inactivation.
Numerical results for both networked SL oscillators and networked R\"ossler oscillators are shown in Fig. \ref{fig:Strategy}.
In both cases, $\rho_{c}$ under strategy 1 is always greater than that under strategy 2, regardless of the heterogeneity/homogeneity of the network topology.
To understand the above results, we plot the curves $\lambda(\rho)$ on the $\lambda$-$\rho$ parameter plane.
As shown in Fig. \ref{fig:Strapanel}, all curves are unimodal with $\lambda(0)$=$\lambda(1)$=0.
Considering the specific processes of the three strategies, it is not difficult to figure out that with the increase of $\rho$, curve 1 increases faster than curve 2 to the maximum, while the decrease from maximum to zero is just the opposite.
Strategies 1 and 2 are two extreme cases, and strategy 3 is between them. As shown in Fig. \ref{fig:Strapanel}, the crosspoints of curves  $\lambda(\rho)$ with the bifurcation curve defined by Eq. (\ref{rhoclambda}) determine that $\rho_c^1>\rho_c^3>\rho_c^2$. Therefore, our theory successfully explain why the low-degree nodes plays more important role than the hubs in terms of dynamical robustness.

\begin{figure*}[!htpb]
  \epsfig{figure=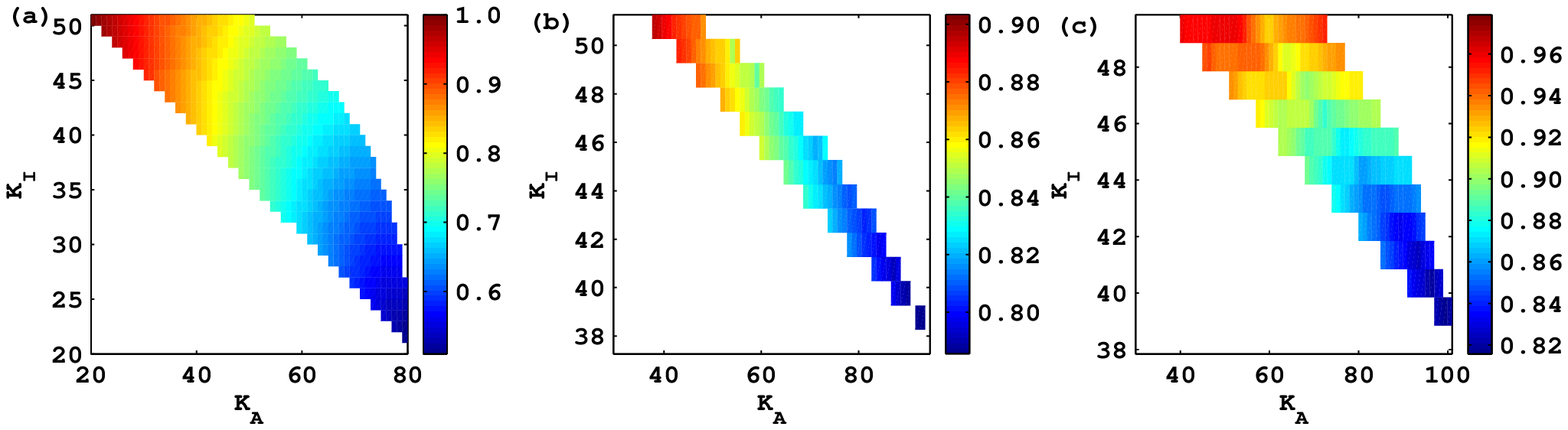,width=1.0\linewidth}
  \caption{(colour online) $\rho_c$ as a function on the panel of $K_A$-$K_I$.  The white area means that $K_A$ and $K_I$ cannot satisfy the constraint Eq. (\ref{eq:kakik}). (a) theoretical result (SL oscillators) with fixed coupling strength and mean degree. (b) Numerical results for networked SL oscillators. (c) Numerical results  for networked R\"ossler oscillators. $ p=0$, $N=500$, $\langle K \rangle=50$.  Other parameters are the same as in Fig. \ref{fig:Strategy}.  }
  \label{fig:kaki}
\end{figure*}

\subsection{Random inactivation}

In the above analysis, the only limitation for the network topology is that it should be dense enough so that the cross link density $\lambda \gg 1$, and there is no requirement for the strategy of inactivation. Thus the above result holds for general inactivation processes. In fact, in our study, we find for typical random inactivation,
Eqs. (\ref{eq:lambdaslaicp1}) and (\ref{eq:lambdaslaicp2}) can take another forms, and the theoretical treatment can be simplified.

We notice that in principle for different oscillator $j$, $K_{j}$ and $\mu_{j}$ are not necessarily the same. However, they must satisfy the constraint $\sum_{j}K_{j}/N=\langle K \rangle$ and $\sum_{j}\mu_{j}/N=\rho$ over all the nodes when the mean degree of the network is fixed.
Let $\mu_{j}=\rho+\xi_{j}$, where $\xi_{j}$ is the deviation of $\mu_{j}$ from its mean value. Normally, a node with larger degree $K_{j}$ would have smaller deviation.  In a dense network, it is reasonable to expect that $|\xi_{j}|\propto 1/K_{j}$ and $\mu_{j}$ distributes symmetrically around its mean value. For random inactivation, the active and inactive oscillators are mixed evenly. Thus the distributions of $\mu_{j}$ in active and inactive subpopulations will approximately the same as in the whole system, i.e., we can approximately assume $\mu_{j}=\rho$ for all $j$.
Then we have $\lambda=L/N=\sum_{j\in S_A}\mu_j K_j/N=\frac{\rho}{N}\sum_{j\in S_A}K_j$, and $\lambda=L/N=\sum_{j\in S_I}(1-\mu_j)K_j/N=\frac{(1-\rho)}{N}\sum_{j\in S_I}K_j$.
Substitute these relations into Eqs. (\ref{eq:lambdaslaicp1}) and (\ref{eq:lambdaslaicp2}), we get
\begin{eqnarray}
\dot{A}=(a+i\Omega-|A|^{2})A+\sigma\rho K_{A}(I-A), \\
\dot{I}=(-b+i\Omega-|I|^{2})I+\sigma(1-\rho)K_{I}(A-I),
\end{eqnarray}
where $K_{A}=\frac{1}{(1-\rho)N}\sum_{j\in S_A}K_j$ and $K_{I}=\frac{1}{\rho N}\sum_{j\in S_I}K_j$ are the mean degrees of active and inactive subpopulations, respectively. Similarly, by applying a linear stability analysis, we can analytically obtain the critical point $\rho_c$ as:
\begin{equation}\label{eq:rhockaki}
\rho_{c}=\frac{ab+\sigma aK_{I}}{\sigma(aK_{I}+bK_{A})}.
\end{equation}
From this equation, we can immediately find that the dynamical robustness of the system are determined by the mean degrees of active and inactive  subpopulations in the case of random inactivation. The point is, even though the mean degree of the whole network is fixed, there is still some degree of freedom for $K_{A}$ and $K_{I}$ as long as they satisfy the following constraint:
\begin{equation}\label{eq:kakik}
\rho K_{I}+(1-\rho)K_{A}=\langle K \rangle.
\end{equation}
On the parameter panel of $K_A$-$K_I$, only certain area can satisfy this condition, as shown in Fig. \ref{fig:kaki}.
In particular, parameters $K_A$ and $K_I$ not only are related to the network topology, but also to the specific strategy of inactivation. Usually, each different realization results in different $K_A$ and $K_I$, causing the small fluctuations of $\rho_c$ observed. We can see in Fig. \ref{fig:kaki} that our theoretical result is qualitatively consistent with the numerical verifications in both networked SL system and R\"ossler system. Extensive numerical results have shown that  Eq. (\ref{eq:rhockaki}) is valid as long as the mean degree is large enough, e.g., $\langle K \rangle \ge 40$.

Furthermore, a trivial solution  always  exists  for  Eq. (\ref{eq:kakik}), i.e., $K_A=K_I=\langle K \rangle$.
In this circumstance, Eq. (\ref{eq:kakik}) degenerates  as:
\begin{equation}\label{eq:K}
\rho_{c}=\frac{a(b+\sigma \langle K \rangle)}{(a+b)\sigma \langle K \rangle}.
\end{equation}
In strictly sense, this result only holds for absolutely homogeneous topologies, i.e., the regular networks. However,
for random inactivation in very homogeneous network, the above conclusion can be expected to hold approximately. In this case the critical point $\rho_c$ only involves the mean degree $\langle K \rangle$ and is almost independent with the strategy of inactivation. Interestingly, this coincides with one situation investigated  in Ref. \cite{Tanaka2012}, where Eq. (\ref{eq:K}) is derived by a different approach. Moreover,
Eqs.  (\ref{eq:rhockaki}) and (\ref{eq:K}) can recover Eq. (4) in Ref. \cite{Daido2004} when the topology is globally connected; and under the strong coupling limit, Eq. (\ref{eq:K}) can reproduce the results in Refs. \cite{pazo2006,Daido2011}. Therefore, all these studies provides insights from different perspectives into the dynamical robustness of such networked system.

\section{Conclusion}

In this work, we have studied the dynamical robustness in a networked system, in which both active and inactive oscillators coexist. With the increase of inactive oscillators in such system, it will gradually lose its global activity. The critical point of the transition characterizes the dynamical robustness of such networked system. Interestingly, we found that the critical point fluctuates even thought the coupling strength and the network topology are fixed. By analytical treatment and numerical experiments, we successfully explain the origin of this phenomenon. Mainly, our study provides the following results: (1) The fluctuation of dynamical robustness in such networked system is caused by the variation of the cross link density. It is the multi-valued dependence of the cross link density on the control parameter, i.e., the ratio of inactive oscillators in the system, that leads to the fluctuation of critical points; (2) Since the inactivation process affects the cross link density more in heterogeneous networks than in homogeneous ones, the fluctuation turns out to be more obvious in the former case; (3) Fluctuations under three typical strategies of inactivation can be predicted based on our theory and verified by numerical simulations. This helps understand the importance of the low-degree nodes in terms of dynamical robustness recently reported in \cite{Tanaka2012}; (4) For random inactivation, the critical point is determined by the mean degrees of subpopulations of both active and inactive oscillators; and in very homogeneous networks, it is only related to the mean degree of the network.
This paper is helpful to understand the collective behaviors of networked dynamical systems, and shed light on certain practical applications as well.

SGG is sponsored by: The Science and Technology Commission of Shanghai Municipality grant No. 10PJ1403300; The Innovation Program of Shanghai Municipal Education Commission grant No. 12ZZ043; and The NSFC grant Nos. 11075056 and 11135001. ZHL is sponsored by The NSFC grant Nos. 10975053 and 11135001.

\appendix
\section{Method to tune the network topology}

Previously, there are several methods to change network topology from homogeneous to heterogeneous \cite{Liu2002,Gardenes2006}. For example,  the degree distribution of network can be tuned between exponential and power-law. However, in the present work, we expect degree distribution to be Poisson form or even Delta function. So we propose a simple but effective method for this purpose. The main idea is to gradually rewind edges from an initially heterogeneous network. Here are the main steps: (1) Generate a heterogeneous network, e.g., the BA network  characterized by a power law degree distribution; (2) Choose an arbitrary edge and compare the degrees of both its ends; (3) Cut the edge from the node with higher degree, and randomly rewind it to a node in the network. We define the rewinding probability $p$ as the number of edges rewound normalized by the total number of edges in the network. When $p$ varies from 0 to 1, the network topology will change from heterogeneous to homogeneous as verified by numerical experiments. Therefore, $p$ is a parameter that can characterize the heterogeneous extent of a network. In our simulations, we have started from scale-free networks with power law exponents between 2 and 3, the results are qualitatively the same.

\end{document}